\documentclass[letter]{aa} 

\usepackage{graphicx}
\usepackage{txfonts,textcomp}
\usepackage{gensymb} 
\usepackage{xspace}
\usepackage{tabularx}
\usepackage{xcolor}
\usepackage{placeins}
\usepackage{threeparttable}
\usepackage{ulem}
\usepackage{natbib,twoopt}
\usepackage[breaklinks=true]{hyperref} 
\bibpunct{(}{)}{;}{a}{}{,} 
\makeatletter
\newcommandtwoopt{\citeads}[3][][]{\href{http://adsabs.harvard.edu/abs/#3}%
{\def\hyper@linkstart##1##2{}%
\let\hyper@linkend\@empty\citealp[#1][#2]{#3}}}
\newcommandtwoopt{\citepads}[3][][]{\href{http://adsabs.harvard.edu/abs/#3}%
{\def\hyper@linkstart##1##2{}
\let\hyper@linkend\@empty\citep[#1][#2]{#3}}}
\newcommandtwoopt{\citetads}[3][][]{\href{http://adsabs.harvard.edu/abs/#3}%
{\def\hyper@linkstart##1##2{}
\let\hyper@linkend\@empty\citet[#1][#2]{#3}}}
\newcommandtwoopt{\citeyearads}[3][][]%
{\href{http://adsabs.harvard.edu/abs/#3}
{\def\hyper@linkstart##1##2{}%
\let\hyper@linkend\@empty\citeyear[#1][#2]{#3}}}
\makeatother
\def\ms{\hbox{m\,s$^{-1}$}}         
\def\m2s2{\hbox{\,m$^{2}$\,s$^{-2}$}} 
\def\Msun{$M_{\odot}$\xspace}             
\def\Rsun{$R_{\odot}$\xspace}
\def\Mjup{\hbox{$\mathrm{M}_{\rm J}$}\xspace}
\def\Rjup{\hbox{$\mathrm{R}_{\rm J}$}\xspace}

\def\ten[#1]{$\;\times 10^{#1}$}

\def\logg{$\log g$}

\newcommand{\e}[1]{{\times10^{#1}}}

\newcommand{\Rnom}{\hbox{$\mathcal{R}^{\rm N}_{\odot}$}} 

\newcommand{\GMnom}{\hbox{$\mathcal{(GM)}^{\rm N}_{\odot}$}}

\newcommand{\Renom}{\hbox{$\mathcal{R}^{\rm N}_{e \rm E}$}}

\newcommand{\GMenom}{\hbox{$\mathcal{(GM)}^{\rm N}_{\rm E}$}}

\newcommand{\RJnom}{\hbox{$\mathcal{R}^{\rm N}_{e \rm J}$}}

\newcommand{\GMJnom}{\hbox{$\mathcal{(GM)}^{\rm N}_{\rm J}$}}

\newcommand{\rebound}{{\sc \tt REBOUND}\xspace}
\newcommand{\whf}{{\sc \tt WHFast}\xspace}
\newcommand{\emcee}{{\sc \tt emcee}\xspace}
\newcommand{\juliet}{{\sc \tt juliet}\xspace}
\newcommand{\batman}{{\sc \tt batman}\xspace}

\newcommand{\celerite}{{\sc \tt celerite}\xspace}

\newcommand{\PARSEC}{{\sc \tt PARSEC}\xspace}

\newcommand{\pipe}{{\sc \tt PIPE}\xspace}
\newcommand{\astropy}{{\sc \tt astropy}\xspace}
\newcommand{\prose}{{\sc \tt prose}\xspace}
\newcommand{\photutils}{{\sc \tt photutils}\xspace}

\newcommand{\Lightkurve}{{\sc \tt Lightkurve}\xspace}

\newcommand{\MEarth}{$\mathrm{M_E}$\xspace}

\newcommand{\be}{\begin{equation}}
\newcommand{\ee}{\end{equation}}

\newcommand{\bea}{\begin{eqnarray}}
\newcommand{\eea}{\end{eqnarray}}

\def\logg{$\log g$}
\def\Msun{$M_{\odot}$\xspace}            
\def\Rsun{$R_{\odot}$\xspace}
\newcommand{\orcid}[1]{\protect\href{https://orcid.org/#1}{\protect\includegraphics[width=8pt]{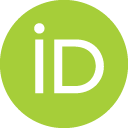}}}
\hypersetup{colorlinks=true, citecolor=blue, linkcolor=blue}

\begin{document} 

   \title{TOI-7510: A solar-analog system of three transiting giant planets near a Laplace resonance chain\thanks{This study uses CHEOPS data observed as part of the Discretionary Programme (DP) PR450028 (PI Almenara).}}
    \titlerunning{TOI-7510}

   \author{
        J.M.~Almenara\orcid{0000-0003-3208-9815}\inst{\ref{Geneva}}
        \and A.~Leleu\orcid{0000-0003-2051-7974}\inst{\ref{Geneva}} 
        \and T.~Guillot\orcid{0000-0002-7188-8428}\inst{\ref{Nice}} 
        \and R.~Mardling\orcid{0000-0001-7362-3311}\inst{\ref{Monash},\ref{Geneva}} 
        \and F.~Bouchy\orcid{0000-0002-7613-393X}\inst{\ref{Geneva}} 
        \and T.~Forveille\orcid{0000-0003-0536-4607}\inst{\ref{Grenoble}}
        \and J.~Winn\orcid{0000-0002-4265-047X}\inst{\ref{Princeton}}
        \and L.~Abe\orcid{0000-0002-0856-4527}\inst{\ref{Nice}}
        \and M.~Beltrame\inst{\ref{Concordia}}
        \and P.~Bendjoya\orcid{0000-0002-4278-1437}\inst{\ref{Nice}}
        \and X.~Bonfils\orcid{0000-0001-9003-8894}\inst{\ref{Grenoble}}
        \and A.~Deline\inst{\ref{Geneva}} 
        \and J.-B.~Delisle\orcid{0000-0001-5844-9888}\inst{\ref{Geneva}}
        \and R.F.~D\'{i}az\orcid{0000-0001-9289-5160}\inst{\ref{ITBA},\ref{BA}}
        \and E.~Frid\'en\orcid{0009-0002-2548-4948}\inst{\ref{Geneva}} 
        \and M.~Hobson\orcid{0000-0002-5945-7975}\inst{\ref{Geneva}}
        \and R.M.~Hoogenboom\orcid{0009-0004-9519-9143}\inst{\ref{Geneva}} 
        \and J.M.~Jenkins\orcid{0000-0002-4715-9460}\inst{\ref{AMES}}
        \and J.~Korth\orcid{0000-0002-0076-6239}\inst{\ref{Geneva}}
        \and M.~Lendl\orcid{0000-0001-9699-1459}\inst{\ref{Geneva}} 
        \and D.~M\'{e}karnia\orcid{0000-0001-5000-7292}\inst{\ref{Nice}}
        \and A.C.~Petit\orcid{0000-0003-1970-1790}\inst{\ref{Nice}}
        \and M.~Rosenqvist\orcid{0009-0009-6885-7450}\inst{\ref{Geneva}}
        \and O.~Su\'{a}rez\orcid{0000-0002-3503-3617}\inst{\ref{Nice}}
        \and A.H.M.J.~Triaud\orcid{0000-0002-5510-8751}\inst{\ref{Birmingham}}
        \and S.~Udry\orcid{0000-0001-7576-6236}\inst{\ref{Geneva}}        
        }
   \institute{
        Observatoire de Gen\`eve, Département d’Astronomie, Universit\'e de Gen\`eve, Chemin Pegasi 51b, 1290 Versoix, Switzerland\label{Geneva}
        \and Universit\'e C\^ote d'Azur, Laboratoire Lagrange, OCA, CNRS UMR 7293, Nice, France\label{Nice}
        \and School of Physics and Astronomy, Monash University, Victoria, 3800, Australia\label{Monash}
        \and Univ. Grenoble Alpes, CNRS, IPAG, F-38000 Grenoble, France\label{Grenoble}
        \and Department of Astrophysical Sciences, Princeton University, NJ 08544, USA\label{Princeton}
        \and PNRA and IPEV, Concordia Station, Antarctica\label{Concordia}
        \and Instituto Tecnol\'ogico de Buenos Aires (ITBA), Iguaz\'u 341, Buenos Aires, CABA C1437, Argentina\label{ITBA}
        \and Instituto de Ciencias F\'isicas (ICIFI; CONICET), ECyT-UNSAM, Buenos Aires, Argentina\label{BA}
        \and NASA Ames Research Center, Moffett Field, CA 94035, USA\label{AMES}
        \and School of Physics \& Astronomy, University of Birmingham, Edgbaston, Birmingham B15 2TT, UK\label{Birmingham}
             }

   \date{}
 
  \abstract
  {
We report the confirmation and initial characterization of a compact and dynamically rich multiple giant planet system orbiting the solar analog TOI-7510. The system was recently identified as a candidate two-planet system in a machine-learning search of the TESS light curves. Using TESS data and photometric follow-up observations with ASTEP, CHEOPS, and EulerCam, we show that one transit was initially misattributed and that the system consists of three transiting giant planets with orbital periods of 11.5, 22.6, and 48.9~days. The planets have radii of 0.65, 0.96, and 0.94~\Rjup, making them the largest known trio of transiting planets. The system architecture lies near a 4:2:1 mean motion resonant chain, inducing large transit timing variations for all three planets. Photodynamical modeling gives mass estimates of 0.057, 0.41, and 0.60~\Mjup and favors low eccentricities and mutual inclinations. TOI-7510 is an interesting system for investigating the dynamical interactions and formation histories of compact systems of giant planets.
    }
   

   \maketitle
%

\section{Introduction}\label{section:introduction}

\citet{Salinas2025} recently identified the solar-analog star TOI-7510 (TIC\,118798035) as a candidate multiplanet system during a machine learning search of the light curves of the Transiting Exoplanet Survey Satellite \citep[TESS;][]{Ricker2015}. They interpreted the system as containing only two planets. However, the apparent transit-timing variations (TTVs) of the outer planet seemed implausibly large \citep[see Fig.~14 in][]{Salinas2025}. This anomaly drew our attention to the system, so we explored the possibility that one of the observed TESS transits was caused by a third planet.

\begin{figure}
  \centering
  \includegraphics[width=0.48\textwidth]{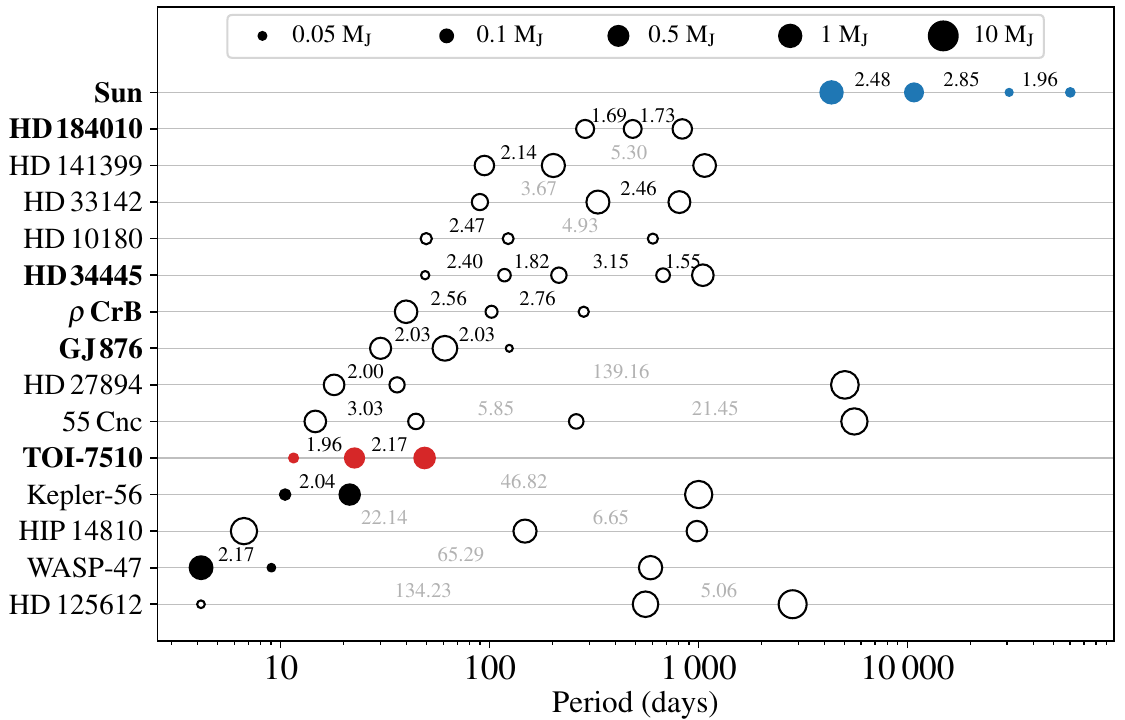}
  \caption{Known systems hosting three or more planets with masses (or minimum masses) between 14~\MEarth and 13~\Mjup, mass uncertainties below 20\%, and orbital period uncertainties below 2\%. Open circles indicate planets detected via radial velocity monitoring, while filled circles correspond to transiting planets. The circle sizes are proportional to the logarithm of the planetary mass. Period ratios between adjacent planets are annotated in black or in light gray if $P_{i+1}/P_i > 3.5$. Notably, the inner pair of TOI-7510 has a period ratio of 1.957, similar to the Uranus–Neptune ratio of 1.961. Lower-mass planets in these systems are excluded from the figure. Data were retrieved from the NASA Exoplanet Archive \citep{Akeson2013}.}\label{figure:periods_3giants}
\end{figure}

In this letter, we confirm three transiting giant planets in the TOI-7510 system with orbital periods of 11.5~days (planet~b), 22.6~days (planet~c), and 48.9~days (planet~d). Their radii are 0.65, 0.96, and 0.94~\Rjup, respectively, making this system the largest known trio of transiting planets in terms of planetary size. Another notable feature is its compactness. There are only five known systems of three or more giant\footnote{Here, we define a giant as a planet \(\geq 14\,M_\oplus\) in order to exclude most of the sub-Neptune population.} planets that feature period ratios of $P_{i+1}/P_i \lesssim 3$,
comparable to the outer Solar System: HD\,184010 \citep{Teng2022}, HD\,34445 \citep{Howard2010,Vogt2017}, $\rho$\,CrB \citep{Noyes1997,Fulton2016,Brewer2023}, GJ\,876 \citep{Delfosse1998,Marcy1998,Marcy2001,Rivera2010}, and now TOI-7510 (see Fig.~\ref{figure:periods_3giants}). Because the four other systems are non-transiting, TOI-7510 is the system of choice for studying the processes that shape compact multi-giant systems such as our own. We highlight this system’s orbital architecture, the preliminary constraints we obtained on its planetary masses and eccentricities, and the high potential of the system for further investigation.

\section{Observations}\label{section:observations}

\begin{figure*}[!ht]
  \centering
  \includegraphics[width=1\textwidth]{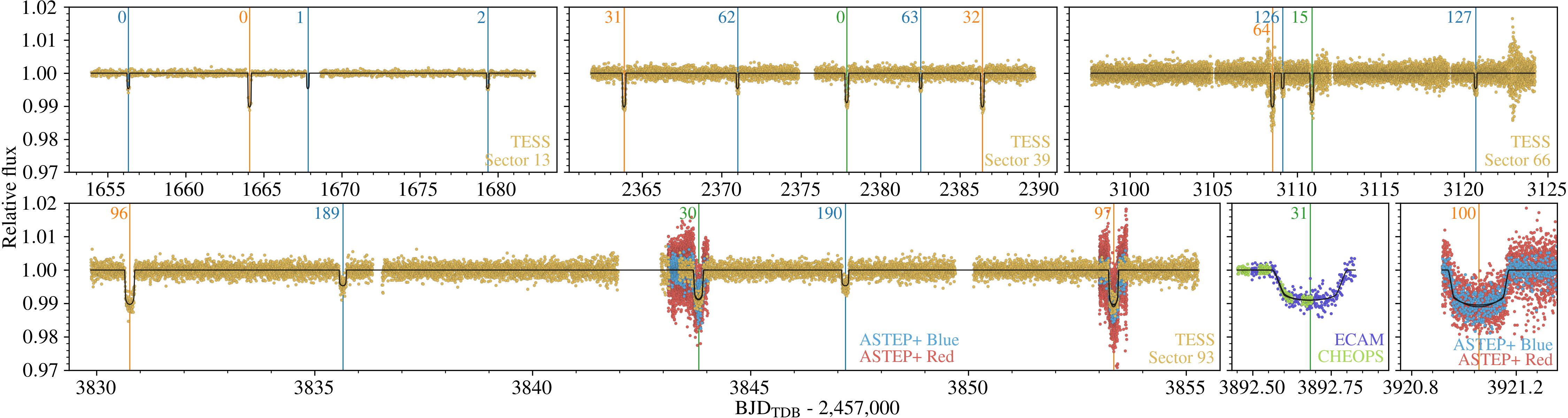}
  \caption{Photodynamical modeling of the transit photometry. The dots, color coded by telescope, represent the noise-model-corrected observations. The black line shows the transit model. Vertical lines mark the midtransit time of planets~b (blue), c (orange), and d (green) and are labeled by the number of orbital periods since the first observed transit.} \label{figure:phot}
\end{figure*}

TOI-7510 was observed by TESS in four sectors, approximately every two years (see Fig.~\ref{figure:phot}). Due to this sparse temporal coverage, the TESS transits do not uniquely constrain the orbital period of planet d, with its possible values ranging from 38.6 to 733 days. To resolve this ambiguity, we organized a follow-up campaign aimed at capturing two consecutive transits of planet~d using the Antarctica Search for Transiting ExoPlanets \citep[ASTEP;][]{Guillot2015,Mekarnia2016}, the CHaracterising ExOPlanet Satellite \citep[CHEOPS;][]{Benz2021}, and the EulerCam \citep[ECAM;][]{Lendl2012}. We confirmed an orbital period of 48.9~days for planet~d. In addition, we acquired 30 radial velocity (RV) measurements with the CORALIE spectrograph \citep{Queloz2001,Segransan2010}. Those datasets are described in Appendix~\ref{section:observations_appendix}.

\section{Stellar parameters}\label{section:stellar_parameters}

To determine the stellar parameters of TOI-7510, we fit its spectral energy distribution (SED; see Appendix~\ref{section:SED}) using stellar atmosphere and evolution models, adopting priors on the distance from the \textit{Gaia} parallax \citep{Gaia, GaiaDR3} and on atmospheric parameters from \textit{Gaia} XP spectra \citep{Andrae2023a}. The derived stellar mass and radius ($1.053 \pm 0.068$~\Msun, $1.030 \pm 0.046$~\Rsun, both within 1~sigma of the Sun) were used as priors in the photodynamical modeling (Sect.~\ref{section:analysis}).

\section{Analysis}\label{section:analysis}

\begin{figure}
  \centering
  \includegraphics[width=0.48\textwidth]{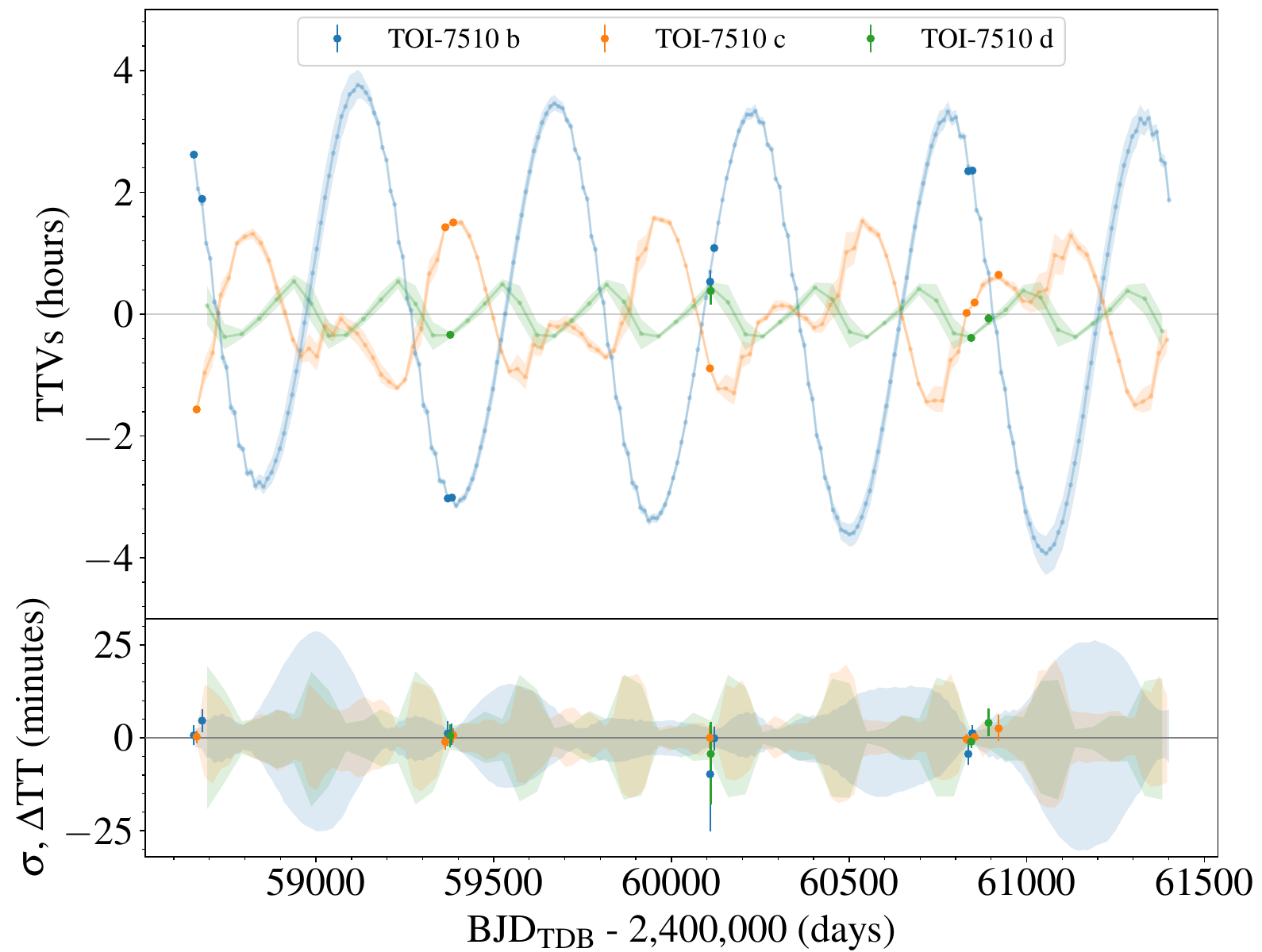}
  \caption{Posterior TTV predictions of planets\,b (blue band), c (orange band), and d (green band) computed relative to a linear ephemeris (Table~\ref{table:results}). A thousand random draws from the posterior distribution were used to estimate the median TTV values and their uncertainties (68.3\% confidence interval). The upper panel shows the posterior TTV values and compares them with the individual transit-time determinations (Table~\ref{table:transit_times}, error bars). In the lower panel, the posterior median transit-timing value was subtracted to emphasize the uncertainty in the distribution and facilitate a clearer comparison with the individually determined transit times.} \label{figure:TTVs}
\end{figure}

We jointly fit the observed photometry and RVs using a photodynamical model \citep{Carter2011} that accounts for the gravitational interactions among the four known bodies in the system (Appendix~\ref{section:photodynamical}). Table~\ref{table:results} lists the median and the 68\% credible interval (CI) of the marginal distribution of the inferred system parameters. 
Figure~\ref{figure:TTVs} presents the posterior TTVs for the three planets, derived from the times of minimum projected separation between the star and planet, and a comparison with the individually determined transit times (Table~\ref{table:transit_times}) computed with \juliet \citep{Espinoza2019,Kreidberg2015,Speagle2020}, assuming a constant transit duration and the same noise model as in the photodynamical analysis.

\section{Results and discussion}\label{section:results}

Pairs of planets near a mean-motion resonance (MMR) are expected to exhibit TTVs at the superperiod, which for a pair of planets near the $k+1:k$ MMR is $P_{super} = 1/(k/P_{in}-(k+1)/P_{out}) $. As the 2:1 MMRs is the closest relevant MMR for both pairs of planets, the two expected superperiods are $P_{super,bc} = 1/(1/P_b-2/P_c) \approx 528$\,days and $P_{super,cd} = 1/(1/P_c-2/P_d) \approx 296 $\,days. Despite the relatively low number of available transits and RV measurements, the large signal to noise ratio of each individual transit observation motivated a photodynamical modeling of the available data (Sect.~\ref{section:analysis}). We note that at this stage, the constraints come mainly from the photometry rather than the RV data due to the limited precision of CORALIE RVs dominated by photon noise for such a star magnitude. As a result, the mass and radius scale of the system is currently set by the stellar mass and radius priors obtained from modeling the star. 
Figure \ref{figure:TTVs} shows that the periodicities of the TTV model for the inner (b) and outer (d) planets are respectively close to the expected values for $P_{super,bc}$ and $P_{super,cd}$, while the TTV model for the middle (c) planet is a sum of contributions at both periods. In addition, a short-term "chopping" effect \citep{Deck2014} is visible in the TTV model. The observed TTVs are therefore well explained by a three-planet model, supporting our interpretation of the system. Regarding the orbital parameters, the orbital inclinations and longitudes of the ascending node are well constrained and compatible with pairwise mutual inclinations of $\lesssim 2^\circ$, hinting at a relatively flat system.

Next we consider the constraints on the masses of the planets and their eccentricity vectors (eccentricity and longitude of periastron). The TTV characterization of multi-planetary systems dominated by a signal at the superperiod can be affected by two layers of eccentricity degeneracies: a degeneracy between the mass of the planets and the part of the eccentricities related to the resonant interaction \citep{Boue2012,Lithwick2012}, and the degeneracy of the non-resonant part of the eccentricities \citep{Leleu2021}.
The first degeneracy has been resolved in tens of TTV systems, enabling robust mass measurements \citep{Hadden2017,Leleu2023,Almenara2025}, typically by observing the chopping signal, the effect of libration in a resonance \citep{Almenara2025}, or other TTV signals besides the superperiod signals. Resolving the second degeneracy, which also allows for the determination of the full eccentricity vector, is more difficult, and it has only been accomplished for a few systems, notably for planets near 2:1 MMRs with giant planets \citep{Korth2023,Korth2024,Borsato2024}. Despite the relatively few observed transits of TOI-7510, the planetary masses appear well constrained, with respective values of $0.057 \pm 0.005$, $0.41 \pm 0.04$, and $0.60 \pm 0.06$~\Mjup for planets b, c, and d. However, given the relatively low number of data points, the mass determinations are not yet robust, and our fit would, for instance, be largely blind to additional dynamically significant planets in the system. The masses may therefore change once more data are obtained. At this point, the full eccentricity vectors of the planets are not well constrained, although their magnitudes are each below 0.1 in the posterior (see Fig.~\ref{figure:ecc}). Overall, the results of this preliminary study are consistent with general trends in the sub-Neptune population, with planets close to MMRs tending to have lower densities (see Fig.~\ref{figure:MR}) and smaller mutual inclinations than than those far from resonance \citep{Leleu2024}.

Regarding the system's architecture, TOI-7510 is the first known system where three transiting giant planets all have radii above 0.65~\Rjup. In fact, two of the planets are nearly Jupiter-sized. Unlike the more common systems of super-Earths and sub-Neptunes, TOI-7510 hosts higher-mass planets with larger envelope masses, suggesting primordial atmospheres and formation during the disk stage. Figure \ref{figure:periods_3giants} presents an analogous comparison involving mass rather than radius, as all other known systems of three or more giants are at least partially non-transiting (open circles in the figure). All three planets of TOI-7510 orbit within 0.27~au, well inside Mercury’s orbital distance around the Sun. The three planets must have undergone substantial inward migration since the inner regions of a typical protoplanetary disk would not contain enough mass to form them in situ.
The period ratios between the planets largely determine the relevance of their mutual gravitational interactions. Only HD\,184010, HD\,34445, $\rho$\,CrB, GJ\,876, and TOI-7510 exhibit compactness comparable to the outer Solar System, with adjacent planets having period ratios $\lesssim3$.
Thus, constraining the current orbital architecture of these systems can lead to better understanding of the physical processes that shaped them in the past, for example, whether the architectures are consistent with formation and migration in the proto-planetary disk and indeed whether they have signatures of post-disk gravitational instabilities \citep[e.g.,][]{Izidoro2017,Li2025}.

Moreover, the period ratios 1.96 and 2.17 for the inner and outer pairs are worth commenting on in this preliminary investigation, especially from the point of view of the formation of a three-planet system composed of a Neptune interior to two gas giants. Studies of planet-disk interactions distinguish between Type I and Type II migration for low- and high-mass planets, respectively, with the latter involving disk clearing in the vicinity of the planet, which in turn results in migration timescales that are considerably longer than for Type I. One can therefore conceive of a situation where planet b arrives at the disk edge ahead of the two giants, and in the meantime, the two giants experience resonant repulsion via mutual density wave emission as they migrate inward \citep{Podlewska-Gaca2012,Baruteau2013,Cui2021}. Since resonance capture normally occurs at period ratios larger than exact commensurability and systems that are not captured move on to the next strong resonance, a period ratio slightly less than commensurability for the inner pair suggests some interesting planet-disk dynamics not readily described by simple migration models. 

With three transiting giant planets near a 4:2:1 Laplace resonance chain with peak-to-peak TTV amplitudes of 1 to 7~hours at the superperiod signal and a typical timing precision of 1 to 3~minutes, TOI-7510 is an excellent candidate for in-depth TTV characterization of its planetary masses and orbital parameters, including the full eccentricity vectors. This should be enabled by the additional TESS observations scheduled for April 21 to June 13, 2026 (sectors 103 and 104), as well as by the future ground- and space-based photometric follow-up. The star is bright enough ($G = 11.8$~mag) for high-precision RV monitoring on an 8~m class telescope, which will help determine precise absolute masses and radii without needing to rely on stellar evolutionary models \citep{Agol2005,Almenara2015}. 

The planets exhibit low bulk densities and moderate stellar insolation, with equilibrium temperatures between 540 and 880~K. Their co-evolution allows for a refined study of their bulk composition \citep{Havel+2011}, while their moderate equilibrium temperatures imply that inflation from Ohmic dissipation should be limited \citep{Thorngren2016}. With its extremely low density, planet~b is particularly intriguing, and it is another example of a so-called super-puff planet \citep{Masuda2014}, probably with a small core and a highly extended H/He envelope. 
Owing to its low density and moderately high equilibrium temperature, planet~b has a transmission spectroscopy metric \citep{Kempton2018} of $155 \pm 17$, above the suggested threshold for atmospheric follow-up. The characterization of the atmospheres of planets in this intriguing near-resonant system should shed light on planetary formation in general.

\section*{Data availability}
Samples from the posterior, transit times forecast, and datasets are available at \url{https://zenodo.org/records/17662413}.

\begin{acknowledgements}

This paper uses data obtained with the ASTEP telescope, at Concordia Station in Antarctica. ASTEP benefited from the support of the French and Italian polar agencies IPEV and PNRA in the framework of the Concordia station program, from OCA, INSU, Idex UCAJEDI (ANR- 15-IDEX-01) and ESA through the Science Faculty of the European Space Research and Technology Centre (ESTEC).
The Birmingham contribution to ASTEP is supported by the European Union's Horizon 2020 research and innovation programme (grant's agreement n$^{\circ}$ 803193/BEBOP), and from the Science and Technology Facilities Council (STFC; grant n$^\circ$ ST/S00193X/1, and ST/W002582/1). 

This paper includes data collected by the TESS mission. Funding for the TESS mission is provided by the NASA's Science Mission Directorate.

Resources supporting this work were provided by the NASA High-End Computing (HEC) Program through the NASA Advanced Supercomputing (NAS) Division at Ames Research Center for the production of the SPOC data products.

This paper includes data collected by the TESS mission that are publicly available from the Mikulski Archive for Space Telescopes (MAST).

CHEOPS is an ESA mission in partnership with Switzerland with important contributions to the payload and the ground segment from Austria, Belgium, France, Germany, Hungary, Italy, Portugal, Spain, Sweden, and the United Kingdom. The CHEOPS Consortium would like to gratefully acknowledge the support received by all the agencies, offices, universities, and industries involved. Their flexibility and willingness to explore new approaches were essential to the success of this mission. CHEOPS data analysed in this article will be made available in the CHEOPS mission archive (\url{https://cheops.unige.ch/archive_browser/}).

This work is based on observations collected with the CORALIE echelle spectrograph and EulerCam mounted on the 1.2~m Swiss Euler telescope at La Silla Observatory, Chile.

This work has made use of data from the European Space Agency (ESA) mission {\it Gaia} (\url{https://www.cosmos.esa.int/gaia}), processed by the {\it Gaia} Data Processing and Analysis Consortium (DPAC, \url{https://www.cosmos.esa.int/web/gaia/dpac/consortium}). Funding for the DPAC has been provided by national institutions, in particular the institutions participating in the {\it Gaia} Multilateral Agreement.

Simulations in this paper made use of the \rebound code which can be downloaded freely at \url{http://github.com/hannorein/rebound}. 

These simulations have been run on the {\it Bonsai} cluster kindly provided by the Observatoire de Gen\`eve.

A.L., J.K, and J.M.A. acknowledges support of the Swiss National Science Foundation under grant number TMSGI2\_211697.
M.L. acknowledges support of the Swiss National Science Foundation under grant number PCEFP2\_194576.

This work has been carried out within the framework of the NCCR PlanetS supported by the Swiss National Science Foundation under grant 51NF40\_205606.

\end{acknowledgements}

\bibliographystyle{aa}
\bibliography{TOI-7510}

\begin{appendix}

\FloatBarrier

\section{Datasets}\label{section:observations_appendix}

\subsection{TESS}

TESS \citep{Ricker2015} observed TOI-7510 in full-frame images (FFIs) during Sectors 13, 39, 66, and 93, with cadences of 1800, 600, 200 and 200 seconds, respectively. For Sectors 13 and 39, we used the presearch data-conditioning simple aperture photometry (PDCSAP; \citealt{Smith2012}, \citealt{Stumpe2012,Stumpe2014}) light curves of TOI-7510, produced by the TESS Science Processing Operations Center \citep[SPOC;][]{Jenkins2016,Caldwell2020}. For sectors 66 and 93, we produced light curves using {\it TESSCut} \citep{Brasseur2019} from FFIs calibrated by SPOC, and performed photometry with the \Lightkurve package \citep{Lightkurve2018}\footnote{BJD$_{\rm TDB}$ times for the target observations were computed following \url{https://github.com/havijw/tess-time-correction}.}. Prior to the availability of SPOC-calibrated FFIs for Sector 93, we also made use of the TESS Image CAlibrator Full Frame Images \citep[TICA;][]{Fausnaugh2020}. 
The SPOC detected the transit signature of planet~c at an orbital period of 22.572~days in a search of Sector~39, and second candidate transit signal at 4.666~days consisting of one transit of planet~b with the single transit of planet~d. Planet~c’s signature passed all the data validation diagnostic tests \citep{Twicken2018}, while the diagnostic tests for the planet b/d transits were compromised by the erroneous orbital period. 
TESS photometry is contaminated by the nearby eclipsing binary Gaia~DR3~6652375098458686720, which has a period of 1.36~days \citep{Mowlavi2023}.

\subsection{ASTEP}

ASTEP is a 0.4 m telescope equipped with a Wynne Newtonian coma corrector, located at Dome C on the east Antarctic plateau \citep{Guillot2015,Mekarnia2016}. The focal box hosts two high-sensitivity cameras at red and blue wavelengths \citep{Schmider2022}. ASTEP observations of TOI-7510 were scheduled on June 16 and 17, 2025, continuously, to attempt to detect planet d whose period, between 38.6 and 733 days, was very uncertain at this point. TESS was observing the target during this period, but a programmed 5-hour data downlink meant the possibility of missing the transit. The transit of planet~d on 2025-06-17 was observed simultaneously and with consistent depths, with ASTEP in band B and band R and TESS. This helped refine the set of 19 possible aliases for the next transit of planet d. Four of these aliases (\#19, 18, 17, and 16) were observed by ASTEP on July 25, 28, and 30, 2025, and August 1, 2025, and ruled out. On August 5, 2025, because of the shorter night, ASTEP could only observe after the predicted egress of alias \#15, whose transit was detected by CHEOPS and EulerCam (see below). In addition, ASTEP observed achromatic transits of planet~c on June 26 and September 2, 2025.

\subsection{CHEOPS}

 CHEOPS \citep{Benz2021} observed a partial transit of planet~d on August 4, 2025 (Table~\ref{table:cheops}). The raw data was automatically processed by the CHEOPS data reduction pipeline \citep[DRP; version 15.1.1;][]{Hoyer2020}. We performed point-spread function photometry with the \pipe package\footnote{\url{https://github.com/alphapsa/PIPE}} \citep{Brandeker2024}.

\begin{table*}
\small
\caption{ CHEOPS observations.}             
\label{table:cheops}      
\centering                          
\begin{tabular}{l c c c c c}        
\hline\hline                 
File key & OBS ID & UTC start & Visit duration & Exposure time & Efficiency \\    
\hline                        
\verb|CH_PR450028_TG000201_V0300|           & 2834211 & 2025-08-04T21:57:43 & 23291 s & 1 x 60.0 s & 60.6\% \\
\verb|CH_PR450028_TG000301_V0300|$^\dagger$ & 2833909 & 2025-08-08T15:32:43 & 26592 s & 1 x 60.0 s & 65.3\% \\
\hline                                   
\end{tabular}
\tablefoot{$^\dagger$The second observation, targeting a different alias of planet~d, had already been scheduled on the satellite when the transit was discovered in the first and could not be canceled.}
\end{table*}

\subsection{EulerCam}

We observed a transit of planet~d, simultaneously with CHEOPS, on August 4, 2025, using EulerCam \citep[ECAM;][]{Lendl2012} on the Swiss 1.2~m Euler telescope at La Silla Observatory. The night was cloudy, and the egress occurred at high airmass. Observations were conducted with the r'-Gunn filter, using exposure times ranging from 30 to 90 seconds. The image reduction was carried out following \citet{Lendl2012} and \citet{Lendl2014}. Aperture photometry was performed using \prose \citep{Garcia2022}, which relies on \astropy \citep{astropy2022} and \photutils \citep{Bradley2023}. The optimal differential photometry followed \citet{Broeg2005}.

\subsection{CORALIE}

TOI-7510 was observed with the CORALIE spectrograph \citep{Queloz2001,Segransan2010} on the Swiss 1.2 m Euler telescope at La Silla Observatory, Chile, between 2025 June 5 and September 14. CORALIE is a fiber-fed echelle spectrograph with a 2\arcsec{} science fiber, and a secondary fiber with a Fabry–Perot for simultaneous drift measurement. It has a spectral resolution of R$\sim$60,000. In total, 30 spectra were obtained with a median exposure duration of 45~minutes and a median signal to noise ratio at 550~nm of 11. RVs are extracted using the standard CORALIE DRS-3.8 by cross-correlating the spectra with a binary G2V mask \citep{Baranne1996,Pepe2002}. The resulting RVs, with a median error of 33~{\ms}, are shown in Fig.~\ref{figure:RV} and listed in Table~\ref{table:RVs}. The BIS, FWHM, and other line-profile diagnostics were also computed, as was the H$\alpha$ index for each spectrum to check for possible variation in stellar activity.
The observation on June 11, 2025, was excluded from the analysis because the spectrum was contaminated by the full Moon located 35\degree\ from the target. 

\begin{table}
\scriptsize
    \renewcommand{\arraystretch}{0.98}
\centering
\caption{CORALIE RV measurements.}\label{table:RVs}
\begin{tabular}{lll}
\hline
\hline
Time & RV &  $\pm1~\sigma$ \\
$[{\rm BJD_{TDB}}]$ & [\ms] & [\ms] \\
\hline
2460831.730192 & 13690.8 & 36.2 \\
2460832.704430 & 13666.6 & 37.8 \\
2460834.776818 & 13645.4 & 27.5 \\
2460835.707845 & 13626.6 & 28.1 \\
2460837.887865$^\dagger$ & 13563.4 & 33.2 \\
2460843.655158 & 13692.5 & 44.3 \\
2460849.693573 & 13638.7 & 56.7 \\
2460851.670032 & 13644.7 & 24.0 \\
2460858.832857 & 13594.1 & 22.7 \\
2460860.647302 & 13626.0 & 18.6 \\
2460863.653596 & 13632.5 & 20.7 \\
2460864.573926 & 13668.9 & 17.3 \\
2460874.828194 & 13710.9 & 23.8 \\
2460875.690735 & 13714.7 & 24.9 \\
2460875.719507 & 13721.7 & 18.3 \\
2460877.779850 & 13709.2 & 31.9 \\
2460895.701032 & 13684.9 & 31.3 \\
2460904.599159 & 13588.6 & 40.8 \\
2460912.632903 & 13722.7 & 34.0 \\
2460916.650961 & 13733.1 & 37.9 \\
2460932.529553 & 13691.4 & 26.1 \\
2460935.665483 & 13735.7 & 20.9 \\
2460939.539438 & 13694.8 & 28.4 \\
2460946.595402 & 13630.4 & 25.2 \\
2460950.535416 & 13639.3 & 32.2 \\
2460953.521484 & 13623.0 & 19.8 \\
2460957.549795 & 13675.7 & 16.9 \\
2460961.569257 & 13688.2 & 40.7 \\
2460964.553022 & 13674.9 & 20.0 \\
2460968.555427 & 13689.6 & 20.3 \\
\hline
\end{tabular}
\tablefoot{$^\dagger$Measurement affected by lunar contamination; excluded from the analysis.}
\end{table}

\section{SED}\label{section:SED}

\begin{figure}[t]
  \centering
  \includegraphics[width=0.5\textwidth]{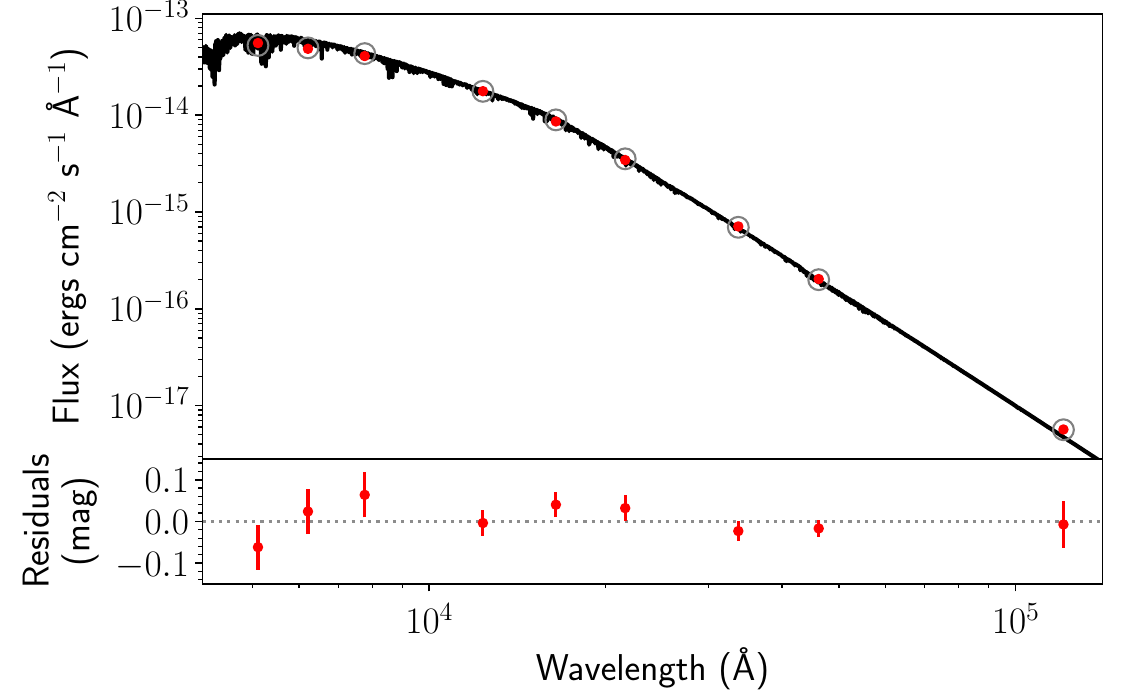}
  \caption{Spectral energy distribution of TOI-7510. The solid line shows the MAP PHOENIX/BT-Settl interpolated synthetic spectrum, red circles are the absolute photometric observations, and gray open circles are the result of integrating the synthetic spectrum in the observed bandpasses. The lower panel shows the residuals of the MAP model, with the jitter added quadratically to the data error bars.} \label{figure:SED}
\end{figure}

The SED was constructed using photometric data from \textit{Gaia} Data Release 3 \citep[DR3;][]{Riello2021}, the 2-Micron All-Sky Survey \citep[2MASS;][]{2mass,Cutri2003}, and the Wide-field Infrared Survey Explorer \citep[WISE;][]{wise,Cutri2013}. The corresponding measurements are listed in Table~\ref{table:stellar_parameters}. 
We adopted the PHOENIX/BT-Settl atmosphere model \citep{Allard2012}, along with two sets of stellar evolution models: Dartmouth \citep{Dotter2008} and \PARSEC \citep{Chen2014}. The SED fitting followed the procedure described by \citet{Diaz2014}, using informative priors for the effective temperature, surface gravity, and metallicity,\footnote{We adopted a solar-scaled composition, assuming $[\rm{\alpha/Fe}]$~=~0.0~dex.} from \citep{Andrae2023b}. The distance prior was based on the \textit{Gaia} zeropoint-corrected parallax \citep{Lindegren2021}. Uniform priors were adopted for the remaining parameters. We included a jitter term \citep{Gregory2005} for each photometric band set (\textit{Gaia}, 2MASS, and WISE). The posterior median values for the stellar radius are identical for both stellar evolution models, while the stellar mass differs by 0.007~\Msun. We merged the results from the two models assuming equal probability for each. The priors and posteriors for all parameters are listed in Table~\ref{table:SED}, and the maximum a posteriori (MAP) stellar atmosphere model is shown in Fig.~\ref{figure:SED}. Following \citet{Tayar2022}, we added a systematic uncertainty floor of 4.2\% for the radius and 5\% for the mass, combined in quadrature with the statistical uncertainties (Table~\ref{table:stellar_parameters}).

\begin{table}[!ht]
    \scriptsize
    \renewcommand{\arraystretch}{1.25}
    \setlength{\tabcolsep}{2pt}
\centering
\caption{Modeling of the SED.}\label{table:SED}
\begin{tabular}{lccc}
\hline
Parameter & & Prior & Posterior median   \\
&  & & and 68.3\% CI \\
\hline
Effective temperature, $T_{\mathrm{eff}}$ & (K)     & $N$(5720, 50)     & $5724 \pm 48$ \\
Surface gravity, \logg\                   & (cgs)   & $N$(4.42, 0.08)   & $4.433 \pm 0.024$ \\
Metallicity, $[\rm{Fe/H}]$          & (dex)   & $N$(0.25, 0.10)   & $0.242 \pm 0.098$ \\
Distance                                  & (pc)    & $N$(245.6, 1.2)   & $245.6 \pm 1.2$ \\
$E_{\mathrm{(B-V)}}$                      & (mag)   & $U$(0, 3)         & $0.062 \pm 0.035$ \\
Jitter \textit{Gaia}                               & (mag)   & $U$(0, 1)         & $0.107^{+0.13}_{-0.050}$ \\
Jitter 2MASS                              & (mag)   & $U$(0, 1)         & $0.035^{+0.056}_{-0.024}$ \\
Jitter WISE                               & (mag)   & $U$(0, 1)         & $0.034^{+0.061}_{-0.025}$ \\
Mass, $M_\star$                           & (M$_\odot$) &               & $1.053 \pm 0.043$ \\
Radius, $R_\star$                         & (R$_\odot$) &               & $1.030 \pm 0.015$ \\
Density, $\rho_\star$                     & ($\mathrm{g\;cm^{-3}}$) &   & $0.965 \pm 0.061$ \\
Isochronal age                            & (Gyr) &                     & $3.0 \pm 2.3$ \\
Luminosity                                & (L$_\odot$) &               & $1.026 \pm 0.040$ \smallskip\\
\hline
\end{tabular}
\tablefoot{$N$($\mu$,$\sigma$): Normal distribution prior with mean $\mu$, and standard deviation $\sigma$. $U$(l,u): Uniform distribution prior in the range [l, u].}
\end{table}

\section{Photodynamical modeling}\label{section:photodynamical}

\begin{figure*}
  \centering
  \includegraphics[width=1\textwidth]{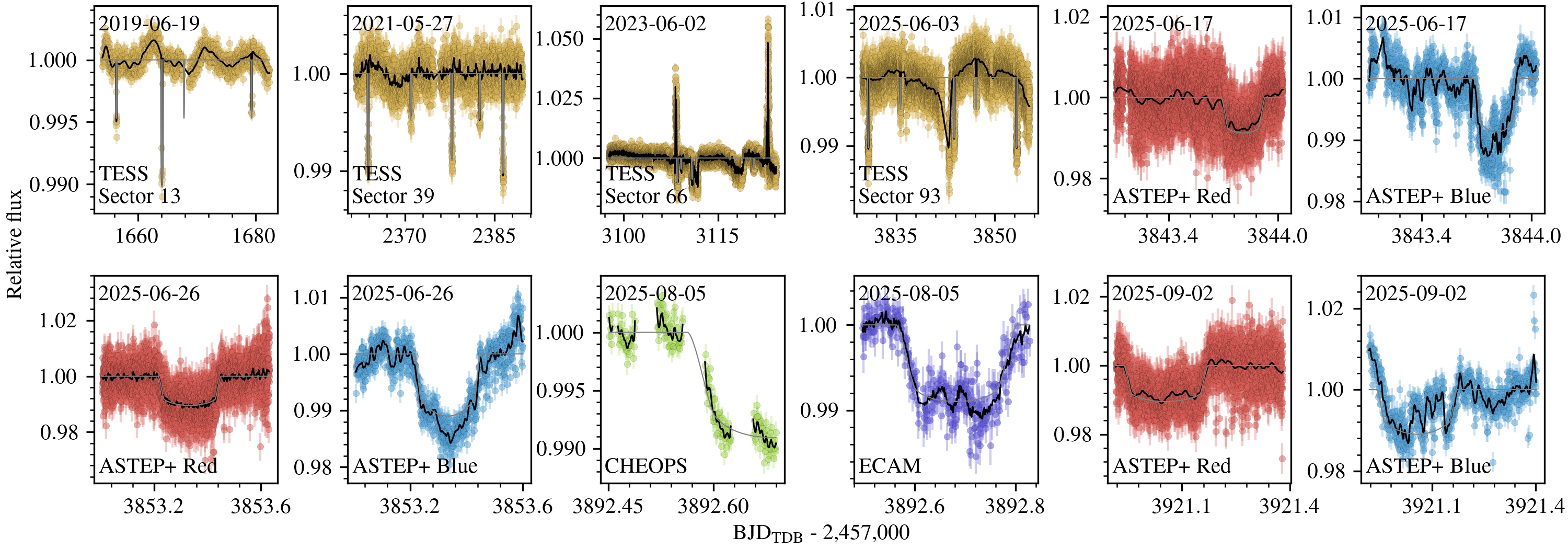}
  \caption{Photodynamical modeling of the transit photometry. Each dataset is shown in a different panel, labeled with the midtransit date (or the start date of the observation for TESS) and the telescope (or instrument). The error bars, in different colors for each telescope (same color code as in Fig.~\ref{figure:phot}), represent the observations. The black line is the MAP model that combines both transits and noise, while the gray line shows the pure transit model.} \label{figure:phot_raw}
\end{figure*}

The positions and velocities of the four bodies assumed in the system were computed as a function of time using the n-body code \rebound \citep{Rein2012} employing the \whf integrator \citep{Rein2015} with a time step of 0.02~days. The sky-projected positions were used to generate the light curve with \batman \citep{Kreidberg2015}, incorporating the light-time effect \citep{Irwin1952}. Oversampling was applied to the TESS data to account for the integration-induced distortion described by \citet{Kipping2010}. The line-of-sight velocity of the star, derived from the n-body integration, was used to model the RV measurements. The model was parameterized using the stellar mass and radius, limb-darkening coefficients, planet-to-star mass and radius ratios, and Jacobi orbital elements (Table~\ref{table:results}) at the reference epoch ($t_{\mathrm{ref}}$). Due to the symmetry of the problem, we fixed the longitude of the ascending node of planet~c at $t_{\mathrm{ref}}$ to 180\degree, and constrained the inclination of planet~d to be less than 90\degree. We modeled the error terms of the transit light curves using Gaussian process regression, adopting the approximate Matern kernel implemented in \celerite \citep{Foreman-Mackey2017}. We used different kernel hyperparameters for each photometry dataset, corresponding to the individual panels in Fig.~\ref{figure:phot_raw}. For the CHEOPS observation, we included a linear model with a $\log$-background term and a sine-of-roll-angle component. For the ECAM transit, we added linear models in the point spread function centroid shifts and the full width at half maximum. The model for the RV has two additional parameters: the systemic velocity and a jitter term. In total, the model comprises 66 free parameters. We adopted normal priors for the stellar mass and radius from Sect.~\ref{section:stellar_parameters}, and uninformative priors for the remaining parameters. The joint posterior distribution was sampled using the \emcee algorithm \citep{Goodman2010, emcee}, with a Gaussian likelihood function and using 660 walkers.  

\begin{figure*}
  \centering
  \includegraphics[width=1\textwidth]{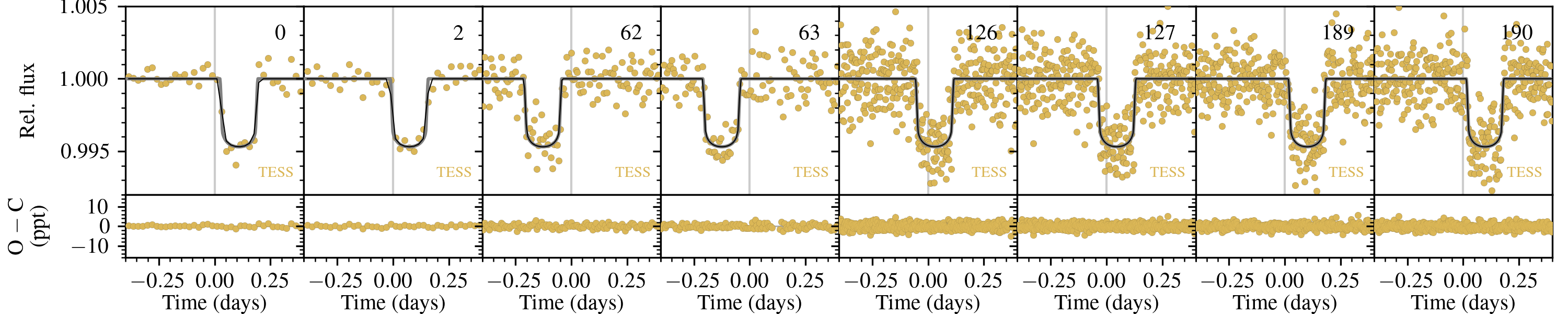}
  \includegraphics[width=1\textwidth]{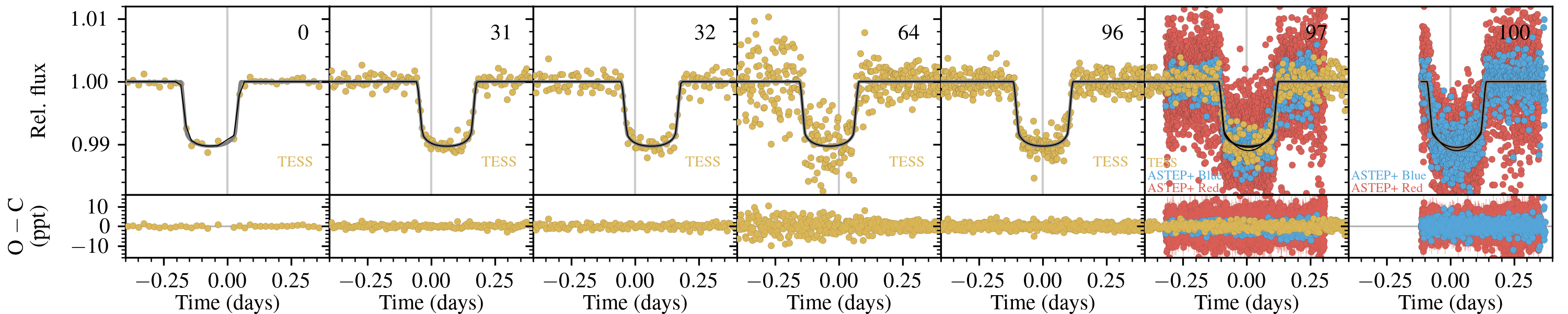}
  \includegraphics[width=1\textwidth]{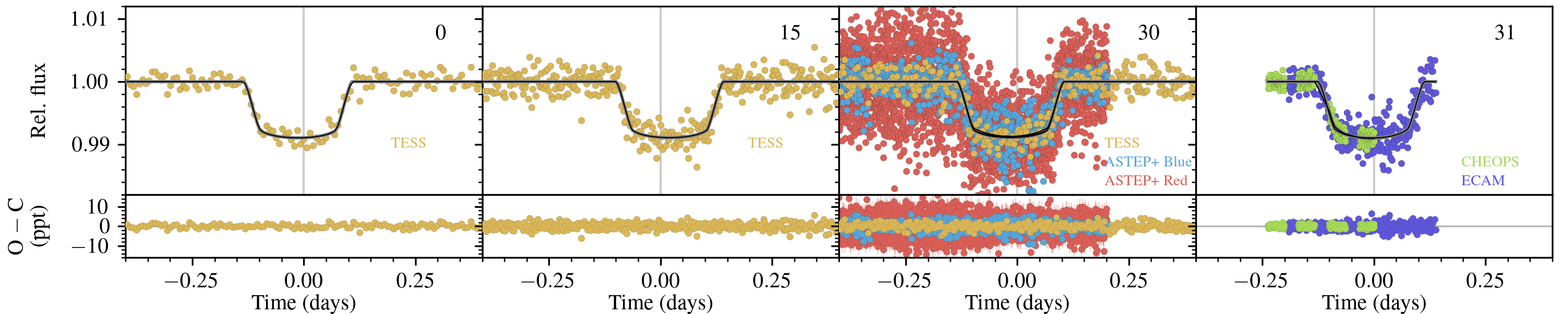}
    \caption{Noise-model-corrected transits of planets b (first row), c (second row), and d (third row), shown as dots with the same color code as in Fig.~\ref{figure:phot}. The MAP model (black line) and the continuous oversampled MAP model (gray line) from the photodynamical modeling are also shown. Each panel is centered at the linear ephemeris (indicated by the vertical gray lines). Each panel is labeled with the telescopes that observed and the epoch number. The residuals after subtracting the MAP model are shown in the lower part of each panel.} \label{figure:transits}
\end{figure*}

\begin{figure}
  \centering
  \includegraphics[width=0.49\textwidth]{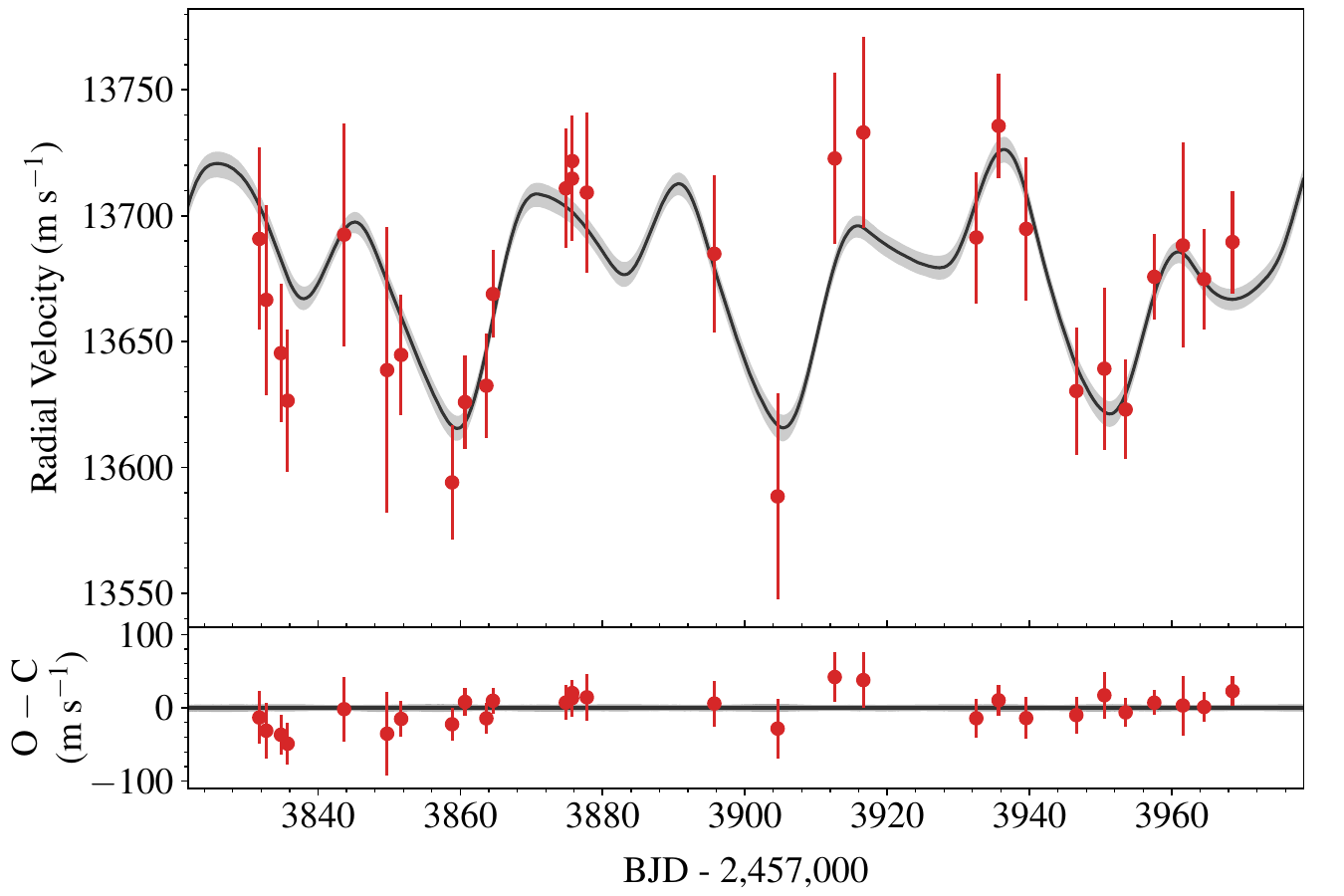}
  \includegraphics[width=0.49\textwidth]{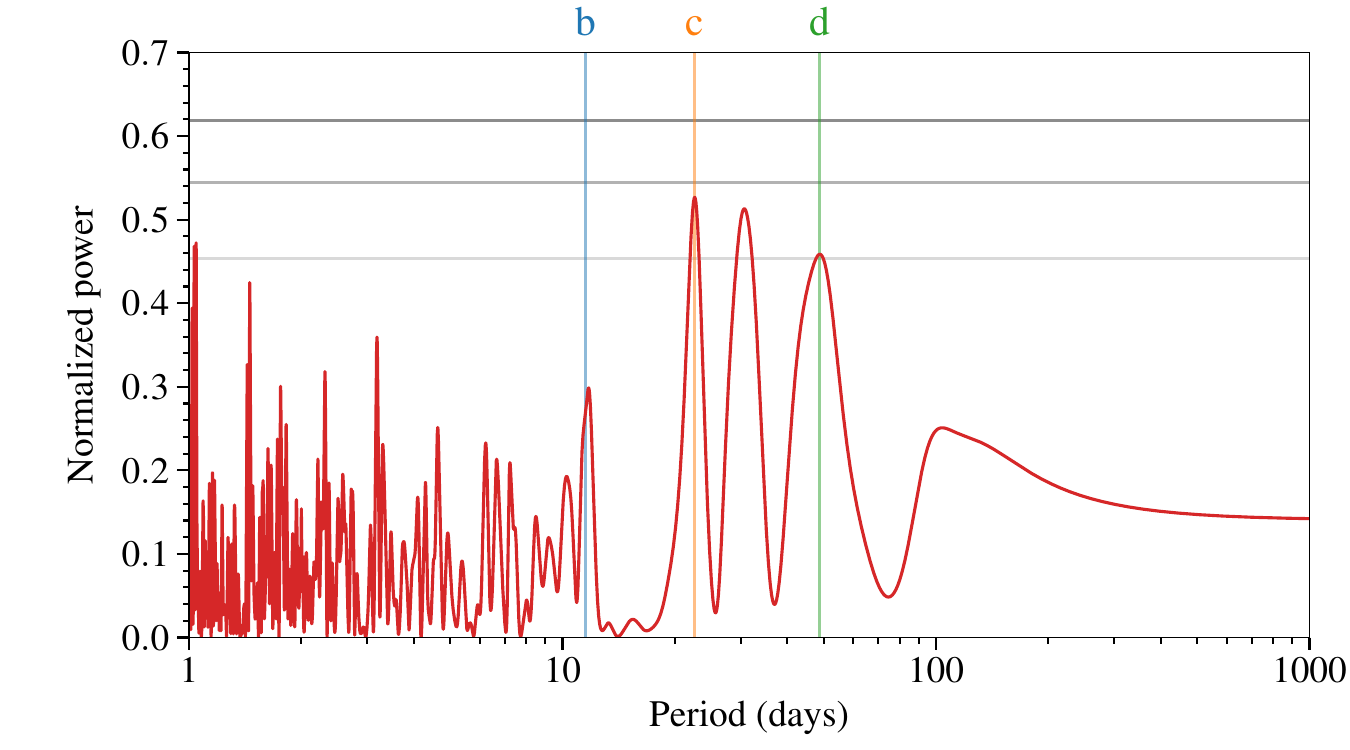}
  \caption{{\it Top panel}: CORALIE RVs of TOI-7510 (red error bars) together with the median RV model (black line) and the 68.3\% confidence interval (gray region), both estimated from a thousand random draws from the posterior distribution. Residuals from the median model are shown. {\it Bottom panel}: Periodogram of the CORALIE RVs. The red line represents the Generalised Lomb–Scargle periodogram \citep[][]{Zechmeister2009}. The gray horizontal lines represent 10, 1, and 0.1\% false-alarm levels, from bottom to top. The vertical blue, orange, and green lines respectively mark the period of the transiting planets b, c, and d.} \label{figure:RV}
\end{figure}

The MAP model for the transit photometry and RV data is plotted in Figs.~\ref{figure:phot_raw}, and \ref{figure:RV}, respectively. Noise-model-corrected datasets and transits for each planet are shown in Figs.~\ref{figure:phot} and \ref{figure:transits}, respectively.

\section{Additional figures and tables}

\begin{table}[htb!]
    \tiny
      \caption{Astrometry, photometry, and stellar parameters for TOI-7510.}\label{table:stellar_parameters}
      \centering
        \setlength{\tabcolsep}{4pt}
    \begin{tabular}{lcl}
    \hline
    \hline
    Parameter & Value & Ref. \\
    \hline
    \textit{Designations} & TYC\,8748-944-1               & 1 \\
                          & 2MASS\,J18145472-5426027      & 2 \\
                          & TIC\,118798035                & 3 \\
                          & Gaia\,DR3\,6652378156475401216 & 4 \medskip\\

    \textit{Astrometry} \\
    Right ascension (ICRS, J2016), $\alpha$ & 18$^{\rm h}$14$^{\rm m}$54.77$^{\rm s}$ & 4 \\
    Declination (ICRS, J2016), $\delta$ & -54$^{\rm o}$26'02.59'' & 4 \\
    Proper motion $\alpha$ (mas/year) & $19.700 \pm 0.017$ & 4 \\
    Proper motion $\delta$ (mas/year) & $6.320 \pm 0.016$ & 4 \\
    Parallax, $\pi$ (mas) & $4.071 \pm 0.020$ & 4, 5 \\
    Distance, d (pc) & $245.6 \pm 1.2$ & $\pi$ \medskip\\
    
    \textit{Photometry} \\
    \textit{\textit{Gaia}}-BP (mag)   & 12.1639 $\pm$ 0.0029  & 4 \\
    \textit{Gaia}-G (mag)    & 11.7832 $\pm$ 0.0028  & 4 \\
    \textit{Gaia}-RP (mag)   & 11.2330 $\pm$ 0.0028  & 4 \\
    2MASS-J (mag)   & 10.593  $\pm$ 0.023  & 2 \\
    2MASS-H (mag)   & 10.275  $\pm$ 0.022 & 2 \\ 
    2MASS-Ks (mag)  & 10.211  $\pm$ 0.023 & 2 \\
    WISE-W1 (mag)   & 10.155  $\pm$ 0.023 & 6 \\
    WISE-W2 (mag)   & 10.189  $\pm$ 0.020 & 6 \\
    WISE-W3 (mag)   & 10.152  $\pm$ 0.056 & 6 \medskip\\
    
    \textit{Stellar parameters} \\
    Spectral type & G3 & 7 \\
    Effective temperature, $T_{\rm eff}$ (K) & $5720 \pm 50$   & 8 \\
    Surface gravity, log g (cgs)             & $4.42 \pm 0.08$ & 8 \\
    Metallicity, [M/H] (dex)                 & $0.25 \pm 0.10$ & 8 \\ 
    Prior stellar mass, $M_\star$ (\Msun)    & $1.053 \pm 0.068$ & 9 \\
    Prior stellar radius, $R_\star$ (\Rsun)  & $1.030 \pm 0.046$ & 9 \\
    \hline
    \end{tabular}
    \begin{tablenotes}
        \tiny
        \item References : 1)~\citet{Hog2000}, 2)~\cite{Cutri2003}, 3)~\citet{Stassun2019}, 4)~\cite{GaiaDR3}, 5)~\cite{Lindegren2021}, 6)~\cite{Cutri2013}, 7)~\cite{Pecaut2013}, 8)~\cite{Andrae2023b}, 9)~This work.
    \end{tablenotes}
\end{table}

\begin{table}[!b]
\centering
    \tiny
    \renewcommand{\arraystretch}{1.25}
    \setlength{\tabcolsep}{2pt}
\caption{Transit times of the observations.}\label{table:transit_times}
\begin{tabular}{rlll}
\hline
Epoch & Posterior median  & Telescope  \\
      & and 68.3\% CI [BJD$_{\mathrm{TDB}}$]  & Instrument  & \\
\hline
\emph{\bf Planet~b}\\ 
0       & $2458656.3228_{-0.0018}^{+0.0019}$ & TESS Sector 13 \\
2       & $2458679.3542_{-0.0021}^{+0.0022}$ & TESS Sector 13 \\
62      & $2459371.0015_{-0.0022}^{+0.0023}$ & TESS Sector 39 \\
63      & $2459382.5329_{-0.0020}^{+0.0021}$ & TESS Sector 39 \\
126     & $2460109.1254_{-0.011}^{+0.0075}$      & TESS Sector 66 \\
127     & $2460120.6792_{-0.0022}^{+0.0021}$ & TESS Sector 66 \\
189     & $2460835.6457 \pm 0.0019$              & TESS Sector 93 \\
190     & $2460847.1770 \pm 0.0015$              & TESS Sector 93 \smallskip\\

\emph{\bf Planet~c}\\ 
0   & $2458664.0919 \pm 0.0012$          & TESS Sector 13 \\    
31  & $2459363.8475 \pm 0.0014$          & TESS Sector 39 \\    
32  & $2459386.4195 \pm 0.0013$          & TESS Sector 39 \\
64  & $2460108.5195_{-0.0020}^{+0.0023}$ & TESS Sector 66 \\    
96  & $2460830.75711 \pm 0.00096$        & TESS Sector 93 \\    
97  & $2460853.33296 \pm 0.00073$        & TESS Sector 93, ASTEP+ Blue \& Red \\  
100 & $2460921.0582_{-0.0023}^{+0.0025}$ & ASTEP+ Blue \& Red \smallskip\\      

\emph{\bf Planet~d}\\ 
0  & $2459377.8765_{-0.0023}^{+0.0022}$ & TESS Sector 39 \\     
15 & $2460110.8734_{-0.0094}^{+0.0060}$ & TESS Sector 66 \\
30 & $2460843.8084_{-0.0012}^{+0.0013}$ & TESS Sector 93, ASTEP+ Blue \& Red \\
31 & $2460892.6861_{-0.0025}^{+0.0027}$ & ECAM, CHEOPS \smallskip\\     

\hline
\end{tabular}
\end{table}

\begin{figure}
  \centering
  \includegraphics[width=0.49\textwidth]{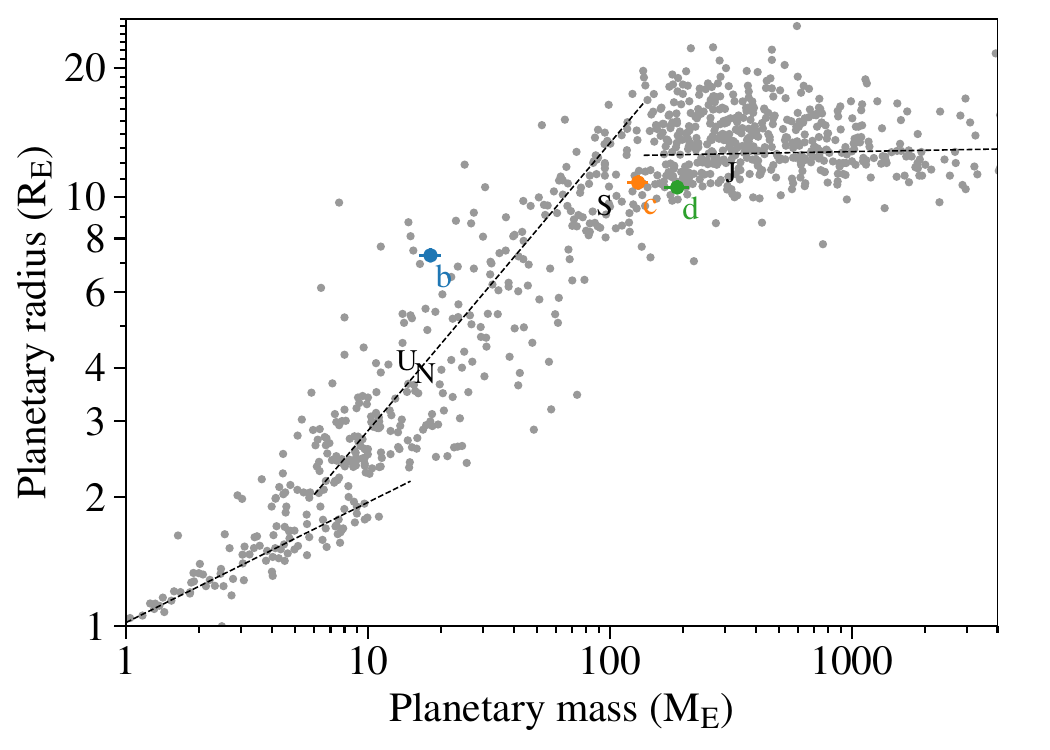}
  \caption{Mass--radius diagram of known exoplanets. Gray dots are transiting planets listed in the NASA Exoplanet Archive with planetary radius and mass uncertainties below 20\%. The dashed black lines are the mass–radius relations from \citet{Parc2024}. The Solar System’s giant planets are indicated by their initials  (J, S, U, N).} \label{figure:MR}
\end{figure}

\begin{figure}
  \centering
  \includegraphics[width=0.47\textwidth]{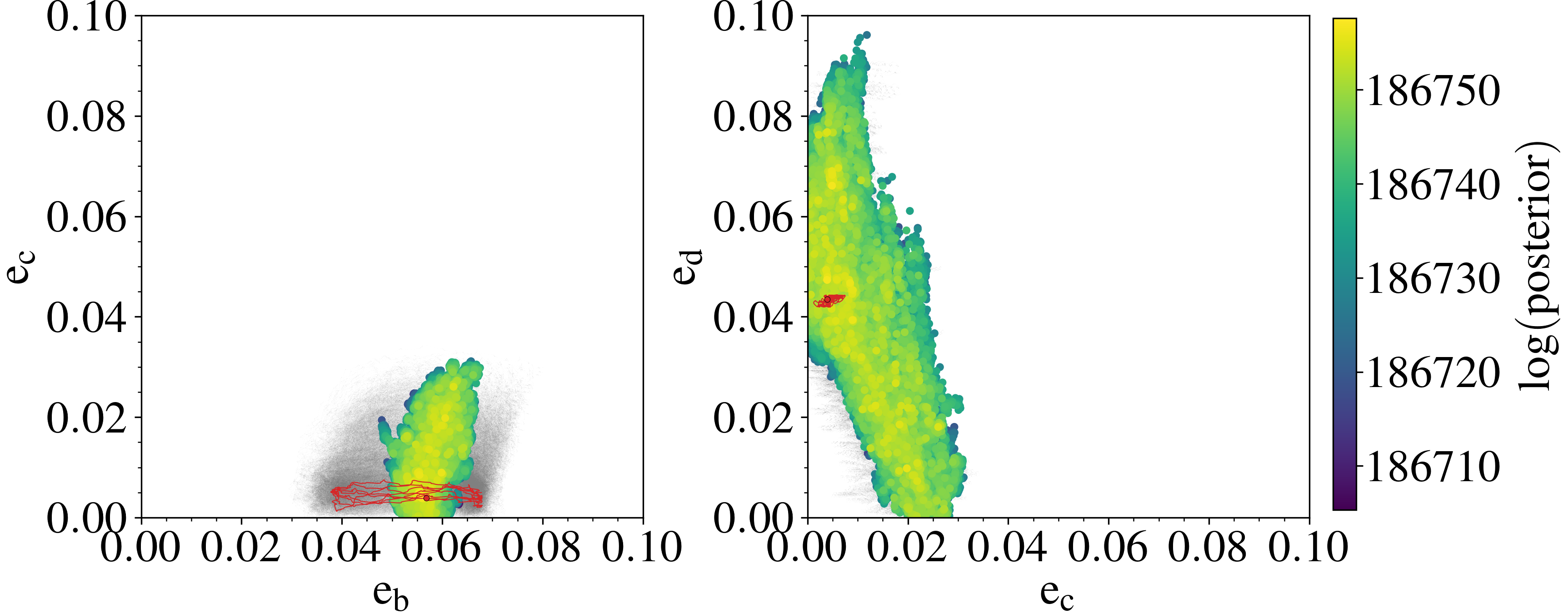}
  \caption{Pairwise joint posterior distributions of the eccentricities for planets~b, c, and d. The dots represent the posterior samples at $t_{\mathrm{ref}}$, with the color scale showing their $\log$-posterior value. The evolution during the observations for a thousand random draws from the posterior distribution is shown in light gray (red for the MAP model).} \label{figure:ecc}
\end{figure}

\begin{table*}
  \small
\renewcommand{\arraystretch}{1.1}
\setlength{\tabcolsep}{4pt}
\centering
\caption{Inferred system parameters.}\label{table:results}
\begin{tabular}{lccccc}
\hline
Parameter & Units & Prior & Median and 68.3\% CI &  &  \\
\hline
\emph{\bf Star} \\
Stellar mass, $M_\star$              & (\Msun)     & $N(1.053, 0.068)$   & $1.063 \pm 0.065$ \\
Stellar radius, $R_\star$            & (\Rnom)     & $N(1.030, 0.046)$   & $1.035 \pm 0.026$ \\
Stellar mean density, $\rho_{\star}$ & ($\mathrm{g\;cm^{-3}}$) &         & $1.372^{+0.050}_{-0.10}$ \\
Surface gravity, \logg\              & (cgs)       &                     & $4.439^{+0.014}_{-0.027}$ \\

\emph{\bf Planets} & &  & \emph{{\bf Planet~b}} & \emph{{\bf Planet~c}} & \emph{{\bf Planet~d}} \\
Semi-major axis, $a$                      & (au)                   &                                           & $0.1019 \pm 0.0020$         & $0.1595 \pm 0.0032$           & $0.2672 \pm 0.0053$        \\
Eccentricity, $e$                         &                        &                                           & $0.0572 \pm 0.0033$         & $0.0091^{+0.010}_{-0.0062}$   & $0.040 \pm 0.022$          \\
Argument of periastron, $\omega$          & (\degree)              &                                           & $95.9 \pm 5.6$              & $187^{+51}_{-21}$             & $338 \pm 37$               \\
Inclination, $i$                          & (\degree)              & $S(0, 180)_{\rm b,c}$, $S(0, 90)_{\rm d}$ & $90.03 \pm 0.72$            & $90.11 \pm 0.34$              & $89.312^{+0.026}_{-0.043}$ \\
Longitude of the ascending node, $\Omega$ & (\degree)              & $U(90, 270)_{\rm b,d}$                    & $180.0 \pm 1.8$             & 180 (fixed at $t_{\mathrm{ref}}$) & $179.63 \pm 0.42$      \\
Mean anomaly, $M_0$                       & (\degree)              &                                           & $116.2 \pm 5.0$             & $263^{+20}_{-63}$             & $170 \pm 39$               \\
$\sqrt{e}\cos{\omega}$                    &                        & $U(-1, 1)$                                & $-0.025^{+0.017}_{-0.024}$  & $-0.087^{+0.074}_{-0.049}$    & $0.155 \pm 0.081$          \\
$\sqrt{e}\sin{\omega}$                    &                        & $U(-1, 1)$                                & $0.2372 \pm 0.0052$         & $-0.011 \pm 0.040$            & $-0.073 \pm 0.098$         \\
Mass ratio, $M_{\mathrm{p}}/M_\star$      &                        & $U(0, 1)$                                 & $(5.12 \pm 0.36)\e{-5}$     & $(36.9 \pm 2.3)\e{-5}$        & $(53.5 \pm 4.6)\e{-5}$     \\
Radius ratio, $R_{\mathrm{p}}/R_\star$    &                        & $U(0, 1)$                                 & $0.06480 \pm 0.00098$       & $0.09578 \pm 0.00092$         & $0.0933 \pm 0.0011$        \\
Scaled semi-major axis, $a/R_{\star}$     &                        &                                           & $21.27^{+0.25}_{-0.55}$     & $33.30^{+0.40}_{-0.85}$       & $55.78^{+0.66}_{-1.4}$     \\
Impact parameter, $b$                     &                        &                                           & $0.18 \pm 0.13$             & $0.14 \pm 0.13$               & $0.681 \pm 0.026$          \\
$T_0'$\;-\;2\;450\;000                    & (BJD$_{\mathrm{TDB}}$) & $U(8386, 10386)$                          & $9382.5322 \pm 0.0016$      & $9386.4191 \pm 0.0011$        & $9377.8713 \pm 0.0021$     \\
$P'$                                      & (d)                    & $U(0, 1000)$                              & $11.52195 \pm 0.00090$      & $22.5617 \pm 0.0031$          & $48.9176 \pm 0.0047$       \\
$K'$                                      & (\ms)                  &                                           & $4.93 \pm 0.36$             & $28.4 \pm 1.9$                & $31.8 \pm 2.8$             \\
Planet mass, $M_{\mathrm{p}}$             &(\MEarth)               &                                           & $18.1 \pm 1.7$              & $131 \pm 11$                  & $190 \pm 20$               \\
                                          &(\Mjup)                 &                                           & $0.0571 \pm 0.0052$         & $0.412 \pm 0.036$             & $0.597 \pm 0.064$          \\
Planet radius, $R_{\mathrm{p}}$           &(\Renom)                &                                           & $7.31 \pm 0.23$             & $10.81 \pm 0.32$              & $10.52 \pm 0.32$           \\
                                          &(\RJnom)                &                                           & $0.652 \pm 0.020$           & $0.964 \pm 0.029$             & $0.939 \pm 0.029$          \\
Planet mean density, $\rho_{\mathrm{p}}$  &($\mathrm{g\;cm^{-3}}$) &                                           & $0.255 \pm 0.026$           & $0.573 \pm 0.062$             & $0.89 \pm 0.11$            \\
Planet surface gravity, $\log$\,$g_{\mathrm{p}}$ &(cgs)            &                                           & $2.521 \pm 0.039$           & $3.042 \pm 0.043$             & $3.224 \pm 0.049$          \\
Equilibrium temperature, T$_{\rm eq}$     & (K)                    &                                           & $878 \pm 13$                & $702 \pm 10$                  & $542.4 \pm 7.8$  \\

Mutual inclination, $I_{b,c}$             & (\degree)              &   &  $1.37^{+1.4}_{-0.78}$  \\
Mutual inclination, $I_{c,d}$             & (\degree)              &   &  $0.97 \pm 0.39$  \\

\emph{\bf RV} \\
Coralie jitter &           &  $U(0, 10)$          & $0.79 \pm 0.14$ \\
Coralie offset & [\ms]     & $U(-10^{5}, 10^{5})$ & $13675.5 \pm 3.9$  \\

\emph{{\bf Linear ephemerides}} \\
Period                  & (d)                    & & 11.5308690  & 22.5687423  & 48.8644680  \\
$T_0$\;-\;2\;450\;000   & (BJD$_{\mathrm{TDB}}$) & & 8656.213656 & 8664.157045 & 8693.788019 \\

\hline
\end{tabular}
\tablefoot{\tiny The table lists: Prior, posterior median, and 68.3\% credible interval (CI) for the photodynamical analysis (Sect.~\ref{section:photodynamical}). The Jacobi orbital elements are given for the reference time $t_{\mathrm{ref}}=2\,459\,386.419785$~BJD$_{\mathrm{TDB}}$. The planetary equilibrium temperature is computed for zero albedo and full day-night heat redistribution. $P'$ and $T_0'$ should not be confused with the linear ephemeris, and they were only used to reduce the correlations between jump parameters, replacing the semi-major axis and the mean anomaly at $t_{\mathrm{ref}}$. The linear ephemerides are derived from the median posterior transit times spanning the years 2019 to 2026. \\$T'_0 \equiv t_{\mathrm{ref}} - \frac{P'}{2\pi}\left(M_0-E+e\sin{E}\right)$ with $E=2\arctan{\left\{\sqrt{\frac{1-e}{1+e}}\tan{\left[\frac{1}{2}\left(\frac{\pi}{2}-\omega\right)\right]}\right\}}$, $P' \equiv \sqrt{\frac{4\pi^2a^{3}}{\mathcal G M_{\star}}}$, $K' \equiv \frac{M_p \sin{i}}{M_\star^{2/3}\sqrt{1-e^2}}\left(\frac{2 \pi \mathcal G}{P'}\right)^{1/3}$. CODATA 2018: $\mathcal G = 6.674\,30$\ten[-11]$\rm{m^3\,kg^{-1}\,s^{-2}}$. IAU 2012: au = $149\,597\,870\,700$~m$\,$. IAU 2015: \Rnom = 6.957\ten[8]~m, \GMnom = 1.327$\,$124$\,$4\ten[20]~$\rm{m^3\,s^{-2}}$, \Renom~=~6.378$\,$1\ten[6]~m, \GMenom = 3.986$\,$004\ten[14]~$\rm{m^3\,s^{-2}}$, \RJnom~=~7.149$\;$2\ten[7]~m, \GMJnom = 1.266$\;$865$\;$3\ten[17]~$\rm{m^3\;s^{-2}}$. \Msun$ = \GMnom/\mathcal G$, \MEarth = \GMenom/$\mathcal G$, \Mjup = \GMJnom/$\mathcal G$, $k^2$ = \GMnom$\,(86\,400~\rm{s})^2$/$\rm{au}^3$. $N(\mu, \sigma)$: Normal distribution with mean $\mu$ and standard deviation $\sigma$. $U(a, b)$: A uniform distribution defined between a lower $a$ and upper $b$ limit. $S(a, b)$: A sinusoidal distribution defined between a lower $a$ and upper $b$ limit.}
\end{table*}

\end{appendix}
\end{document}